\documentclass[aps,prd,amssymb,twocolumn,10pt]{revtex4-1}
\newcommand*\widebar[1]{%
  \hbox{%
    \vbox{%
      \hrule height 0.5pt 
      \kern0.5ex
      \hbox{%
        \kern-0.1em
        \ensuremath{#1}%
        \kern-0.1em
      }%
    }%
  }%
}
\usepackage{graphicx}
\usepackage{dcolumn}%
\usepackage{bm}%
\usepackage[]{complexity}

\usepackage[version=3]{mhchem}
\usepackage{slashed}

\usepackage[bookmarks=false]{hyperref}

\begin{document}


\title{Charge asymmetry in elastic scattering of massive leptons on protons}


\author{Oleksandr Koshchii}
\email[]{koshchii@gwmail.gwu.edu}
\affiliation{The George Washington University, Washington, D.C. 20052, USA}

\author{Andrei Afanasev}
\email[]{afanas@gwu.edu}
\affiliation{The George Washington University, Washington, D.C. 20052, USA}


\date{\today}

\begin{abstract}
The charge asymmetry due to higher-order QED corrections in elastic lepton-proton scattering is estimated without employing the ultrarelativistic approximation. Our calculation is performed by generalizing the soft-photon approximation approach suggested by Tsai. Corresponding loop integrals that take a form of Passarino-Veltman scalar three-point functions are calculated analytically without neglecting the mass of the lepton. Our results provide model-independent charge asymmetry predictions for scattering of unpolarized and massive leptons on proton targets. These predictions can be used in corresponding experiments to determine the contribution coming from model-dependent hard two-photon exchange processes.
\end{abstract}

\maketitle
\section{Introduction}

Elastic lepton scattering off of a nucleus has proved to be an efficient tool to study the structure of nucleons. Electrons, which are the lightest known charged leptons in nature, are used to the fullest extent from this perspective. Experiments at JLab, MAMI, SLAC, DESY, Novosibirsk, etc., performed unpolarized and/or polarized cross section measurements by scattering elastically ultrarelativistic electrons off hadron targets. As a result of these measurements, the electric $G_E^N (Q^2)$ and magnetic $G_M^N (Q^2)$ form factors of the target nucleus ($N$) can be extracted. These form factors describe charge and magnetization distributions within the nucleus.

Until recently, the Rosenbluth separation technique, which can be applied to a scattering of unpolarized leptons, has been used extensively to study the $Q^2$ dependence of the electric and magnetic form factors. However, this technique has a substantial drawback: it leads to large uncertainties at momentum transfers $Q^2 \gtrsim 1$ GeV$^2$. In this kinematic region, cross section measurements become less sensitive to the electric form factor $G_E(Q^2)$. To improve the accuracy of the Rosenbluth separation technique, the idea to employ a polarization transfer method was suggested in Ref. \cite{Akhiezer_Rekalo_1974}. Instead of measuring electric and magnetic form factors separately, the authors put forward the idea to access the $G_E$-to-$G_M$ ratio by detecting the polarization of the recoil nucleon in elastic scattering of polarized electrons off of unpolarized nucleon targets. The unique feature of this method is that it does not suffer from the dramatically reduced sensitivity to the $G_E$ component.

The first accurate measurements of the ratio $G^p_E/G^p_M$ by employing the double polarization method \cite{PhysRevLett.88.092301, PhysRevLett.84.1398} revealed a considerable discrepancy in the ratio compared to the results of the Rosenbluth separation. This discrepancy, sometimes referred as the proton form factor puzzle, can be explained to a large extent by the presence of ``hard'' two-photon exchange (TPE) corrections \cite{PhysRevLett.91.142304, PhysRevLett.93.122301}. Corresponding theoretical calculations, in contrast to the ``soft'' TPE estimations, take proton structure effects into consideration. Therefore, these computations are model dependent. In Refs. \cite{PhysRevLett.91.142304, PhysRevC.72.034612}, the authors calculated hard TPE using the hadronic framework that is usually justified at $Q^2 \lesssim 1$ GeV$^2$, whereas the authors of Refs. \cite{PhysRevD.72.013008, PhysRevLett.93.122301} estimated TPE based on the partonic model that works well at $Q^2 \gtrsim 1$ GeV$^2$. More recently, TPE corrections were considered using the dispersion relations formalism \cite{Gorchtein2007322, Tomalak2016}. Despite all these significant theoretical efforts being directed at understanding the physics of TPE, for the present, there is no complete calculation valid at all kinematics. The detailed information on the recent progress in studying TPE can be found in reviews \cite{carlson2007two, Arrington2011782, Afanasev:2017gsk, Blunden:2017nby}.

Besides affecting the $G_E^p/G_M^p$ ratio, TPE is expected to play an important role in a precise determination of the charge radius of a nucleon. From this perspective, a genuine interest in a better understanding of TPE effects is drawn by the so-called proton radius puzzle problem \cite{pohl2010size, Antognini417}. According to Ref. \cite{RevModPhys.84.1527}, there is a large ($\sim 7 \sigma$) discrepancy between the electron- and muon-based charge radii of the proton. The electron-based value $r_{ch} = 0.8775(51)$ fm is obtained following from the results of both hydrogen spectroscopy measurements and elastic electron-proton scattering data. The muon-based value $r_{ch} = 0.84087(39)$ fm is obtained following from the results of atomic spectroscopy of a muonic hydrogen. Currently, there exists no precisely determined value for $r_{ch}$ extracted from elastic muon-proton scattering. Supposedly, a better understanding of muonic TPE can shed light on this discrepancy. As it is discussed in Ref. \cite{Carlson201559}, TPE contributes to the energy shift in atomic systems, thereby affecting the radius of the proton extracted from spectroscopy measurements. In addition, TPE needs to be taken into consideration to extract the radius of the proton in future elastic muon-proton scattering measurements.

The proton form factor and the proton radius puzzles have triggered many experimental endeavors aimed at the precise determination of TPE effects. From this perspective, TPE corrections can be directly accessed by studying the asymmetry (or ratio) between elastic $l^+ p$ and $l^- p$ scattering cross sections. It appears that the interference between one- and two-photon exchange amplitudes along with the interference between lepton and proton bremsstrahlung radiation are the only charge-dependent contributions to order $\alpha^3$. Modern facilities have recently achieved a precision that enables us to measure effects of this order. Recent experiments at VEPP-3, OLYMPUS and CLAS \cite{PhysRevLett.114.062005, PhysRevLett.118.092501, Rimal:2016toz}, which determined TPE using the positron-to-electron scattering cross section ratio, provide a valuable input for our understanding of TPE.

The recently performed PRad \cite{gasparian2014prad} and the future MUSE \cite{gilman2013studying} experiments will address the proton radius puzzle and they require radiative corrections analysis to be done on the subpercent level. This means that TPE processes have to be included in their analysis, as well. The distinct feature of MUSE is that, simultaneously with a precise extraction of the radius from elastic $e^\pm p$ scattering data, the radius is going to be measured in elastic $\mu^\pm p$ scattering. Incoming electron and muon beam momenta there are going to be $115, \ 153,$ and $210$ MeV. In this kinematics, an extra theoretical complication comes out. Namely, a widely used ultrarelativistic (UR) approximation cannot be employed in MUSE to describe the scattering of muons. In other words, the mass of the muon is going to be comparable to its energy and thus cannot be neglected. This means that older radiative corrections codes naturally using the UR limit to describe the scattering of ``light'' electrons have to be reconsidered. In particular, not only hard- but also soft-photon contributions have to be revised.

Given the ongoing demand in including lepton mass effects in radiative corrections calculations for elastic $\mu^\pm p$ scattering, as well as expecting these effects to play an important role in precise $e^\pm p$ scattering measurements at low energies, we perform a model-independent $l^\pm p$ charge asymmetry calculation. This calculation is an opening study aimed at modifying an existing radiative corrections formalism to include the mass of the lepton in final expressions. It should be mentioned here that the separation of a photon's phase space into the soft (model-independent) and hard (model-dependent) regions is not unique. The commonly accepted prescriptions are those of Tsai \cite{PhysRev.122.1898} and Maximon and Tjon \cite{PhysRevC.62.054320}. In their paper, Maximon and Tjon removed many of the mathematical approximations of the original work of Tsai. In particular, they calculated exactly the bremsstrahlung interference contribution, whereas the corresponding calculation of Tsai is approximate; for the detailed discussion on the difference between the two approaches, please see Ref. \cite{Gerasimov2015}. Despite many advantages of the work of Maximon and Tjon, we believe that the work of Tsai is more self-consistent in its definition of soft photons. Briefly, in Ref. \cite{PhysRevC.62.054320}, the authors set the momentum of the soft photon to be zero only in numerators of TPE amplitudes, whereas the author of Ref. \cite{PhysRev.122.1898} sets it to be zero in numerators and denominators of corresponding amplitudes. In our opinion, the latter approach is preferable because the momentum of soft photons appears in numerators of those amplitudes due to simultaneous multiplication of numerators and denominators of intermediate particles propagators by a factor that includes the momentum of soft photons. That is why setting the momentum of soft photons in both numerators and denominators of TPE amplitudes seems to be more self-consistent. For this reason, in our paper we decided to follow the soft-photon definition of Ref. \cite{PhysRev.122.1898}. Using this definition, we generalize the entire approach to include the mass of the lepton and perform calculations of soft TPE and bremsstrahlung interference contributions without any additional approximations that were made originally by Tsai.

The organization of the paper is as follows. In Sec. \ref{1.44} we introduce the general lepton-proton scattering formalism and provide corresponding beyond UR approximation Born expressions. In addition, we identify higher-order QED corrections that make a difference in the comparison of elastic $l^+ p$ vs $l^- p$ scattering. In Sec. \ref{1.45} we provide details of the evaluation of two-photon exchange diagrams using the soft-photon approximation. The corresponding TPE contribution is expressed through Passarino-Veltman three-point functions, the analytical form for which is given in Appendix \ref{1.39}. Section \ref{1.46} describes the soft-photon bremsstrahlung contribution calculation required for cancellation of the infrared divergent result of Sec. \ref{1.45}. Section \ref{1.47} provides the analytical result of the charge asymmetry calculation and corresponding predictions for the MUSE experiment. The conclusion is given in Sec. \ref{1.48}.

\section{Lepton-Proton Scattering Formalism}\label{1.44}

In this paper, we calculate the charge asymmetry in the elastic lepton-proton scattering process that schematically can be written as
\begin{equation}\label{1.1}
    l^\pm (k_1) + p(p_1) \rightarrow l^\pm(k_2) + p(p_2),
\end{equation}
where, in the laboratory frame, the following notation is chosen for the 4-momenta of incoming and outgoing particles: $k_1 = (\varepsilon_1, \vec{k}_1)$, $k_2 = (\varepsilon_2, \vec{k}_2)$, $p_1 = (E_1, \vec{p}_1)$, $p_2 = (E_2, \vec{p}_2)$. Due to a finite detector resolution, this process is always supplemented by the indistinguishable radiative process
\begin{equation}\label{1.2}
    l^\pm (k_1) + p(p_1) \rightarrow l^\pm(k_2) + p(p_2) + \gamma(k),
\end{equation}

\begin{figure*}[htp]
    \includegraphics[scale=0.4]{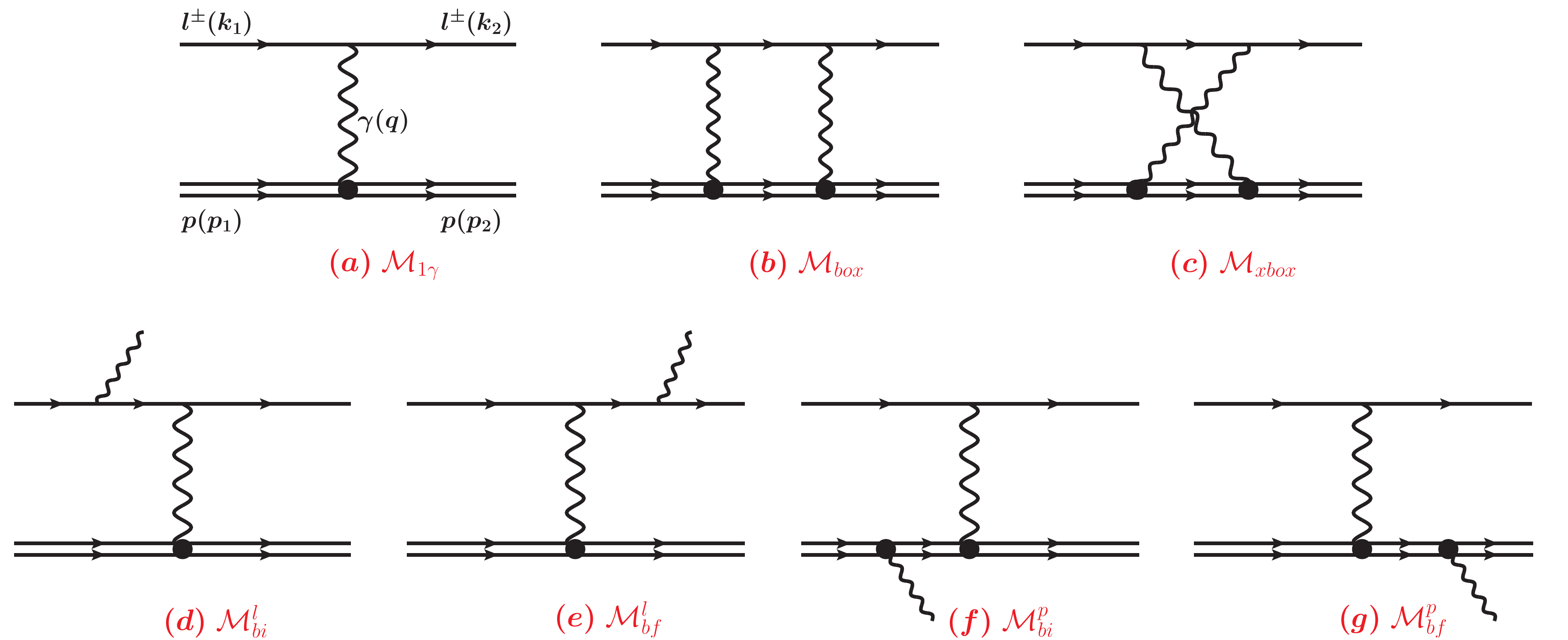}
    \caption{\label{fig:1}Elastic lepton-proton scattering diagrams that contribute to the charge asymmetry to order $\alpha^3$}
\end{figure*}

where the 4-momentum of the emitted photon in the lab frame is given by $k = (\omega, \vec{k})$.

The differential cross section for the unpolarized elastic process (\ref{1.1}), summed over the final and averaged over the initial spin states, can be written as \cite{berestetskii1980quantum}
\begin{equation}\label{1.3}
    {d \sigma} = \frac{1}{(4 \pi)^2} \frac{1}{4 M^2} \frac{\vec{k}_2^2}{|\vec{k}_1| \Big( |\vec{k}_2| + \frac{\varepsilon_1}{M} |\vec{k}_2| - \frac{\varepsilon_2}{M} |\vec{k}_1| \cos \theta \Big)} |\widebar{\mathcal{M}}|^2,
\end{equation}
where $M$ is the rest mass of the proton and $\theta$ is the lab frame scattering angle. The energy $\varepsilon_2$ of the scattered lepton of mass $m$ is given by \cite{Gakh201552}
\begin{equation}\label{1.4}
    \varepsilon_2 = \frac {(\varepsilon_1 + M)(\varepsilon_1 M + m^2) + \vec{k}_1^2 \cos \theta \sqrt{M^2 - m^2 \sin^2 \theta}}{(\varepsilon_1 + M)^2 - \vec{k}_1^2 \cos^2 \theta}.
\end{equation}

Most of our results will be presented in the invariant form, for which we define
\begin{multline}\label{1.5}
    s = (k_1 + p_1)^2, \ \ q^2 = (k_1 - k_2)^2 = - Q^2, \\
    u = (k_1 - p_2)^2, \ \ b_{ij} \equiv 2 \ (k_i \cdot p_j), \ \  i,j = 1,2.
\end{multline}
Here, we should note that, due to the 4-momentum conservation in the elastic scattering process (\ref{1.1}), the following identities are applied there:
\begin{equation}\label{1.6}
    b_{11} = b_{22} \ \ \ \mathrm{and} \ \ \ b_{12} = b_{21}.
\end{equation}

The present calculation focuses on obtaining analytical expressions for the leading-order charge-odd contributions to the scattering amplitude $\mathcal{M}$ of Eq. (\ref{1.3}). These contributions play a difference in the comparison of elastic $l^+$ vs $l^-$ scattering off of the proton, and this difference is measurable in modern experiments. The respective charge-dependent amplitudes, the Feynman diagrams for which are shown in Fig. \ref{fig:1}, can be written as
\begin{multline}\label{1.7}
    |\mathcal{M}|^2 = 2 \mathrm{Re} \Big[ M^{\dagger}_{1 \gamma} \cdot (M_{box} + M_{xbox}) \\
    + (M^l_{bi} + M^l_{bf})^\dagger \cdot (M^p_{bi} + M^p_{bf}) \Big] + \mathcal{O} (\alpha^4).
\end{multline}

In our approach, both terms of Eq. (\ref{1.7}) can be factorized by the square of the one-photon exchange (Born) amplitude $M_{1 \gamma}$, which is given by
\begin{equation}\label{1.8}
    M_{1 \gamma} = z \frac{i e^2}{Q^2} \bar{u}(k_2) \gamma^\mu u(k_1) \bar{U}(p_2) \Gamma_\mu U(p_1),
\end{equation}
where $z = \mp 1$ corresponds to the scattering of $l^\pm$. In addition, the on-shell proton vertex $\Gamma_\mu$ is defined as
\begin{equation}\label{1.9}
    \Gamma_\mu (q) = \gamma_\mu F_1 (q^2) + \frac{i \sigma_{\mu \nu} q^\nu}{2M} F_2 (q^2),
\end{equation}
where $\sigma_{\mu \nu} = \frac{i}{2}[\gamma_\mu, \gamma_\nu]$, and $F_1 (q^2)$ and $F_2 (q^2)$ are the Dirac and Pauli form factors.
The factorization of the scattering amplitude ($|M|^2 \sim |M_{1 \gamma}|^2$) implies that by plugging Eq. (\ref{1.7}) into Eq. (\ref{1.3}) one may obtain the following form for the charge-dependent differential cross section,
\begin{equation}\label{1.10}
    d \sigma^\pm = d \sigma_{1 \gamma} \Big(1 \pm \delta \Big),
\end{equation}
where the asymmetry $\delta$ is defined to be
\begin{equation}\label{1.11}
     \delta \equiv \frac{d \sigma^+ - d \sigma^-}{d \sigma^+ + d \sigma^-}.
\end{equation}
In addition, the Born cross section $d \sigma_{1 \gamma}$ for the scattering of the massive lepton off of the proton target has the following form \cite{PhysRevC.36.2466}:
\begin{multline}\label{1.12}
    d \sigma_{1 \gamma} = \frac{1}{\epsilon_m (1 + \tau)} \Big[ \tau G_M^2(Q^2) + \epsilon_m G_E^2(Q^2) \Big] {d \sigma_M}, \\
    d \sigma_M = \frac{\alpha^2}{Q^4} \frac{\Big( 4 \varepsilon_1 \varepsilon_2 - Q^2 \Big) \vec{k}_2^2}{|\vec{k}_1| \Big( |\vec{k}_2| + \frac{\varepsilon_1}{M} |\vec{k}_2| - \frac{\varepsilon_2}{M} |\vec{k}_1| \cos \theta \Big)},  \\
    \epsilon_m^{- 1} = \frac{(s-u)^2 + Q^2 (4 M^2 + Q^2) - 4 m^2 (4 M^2 + Q^2)}{(s-u)^2 - Q^2 (4 M^2 + Q^2)}.
\end{multline}

It should be noted here that in our paper we call the first term of Eq. (\ref{1.7}) the TPE contribution and the second term of the same equation the bremsstrahlung contribution. Whenever considered independently, these contributions appear to be infrared divergent. The standard approach in dealing with these divergences is to assign an infinitesimally small mass $\lambda$ to the photon, and split the photon's phase space into soft and hard regions. Soft TPE and soft bremsstrahlung contributions have to absorb the infrared-divergent $\lambda$-dependence, whereas respective hard contributions have to be finite. The sum of soft pieces comes out to be $\lambda$ independent and this result provides a physical justification for the theory.

\section{Two-photon exchange box and crossed box contributions}\label{1.45}
\begin{figure}[htp]
    \includegraphics[scale=0.43]{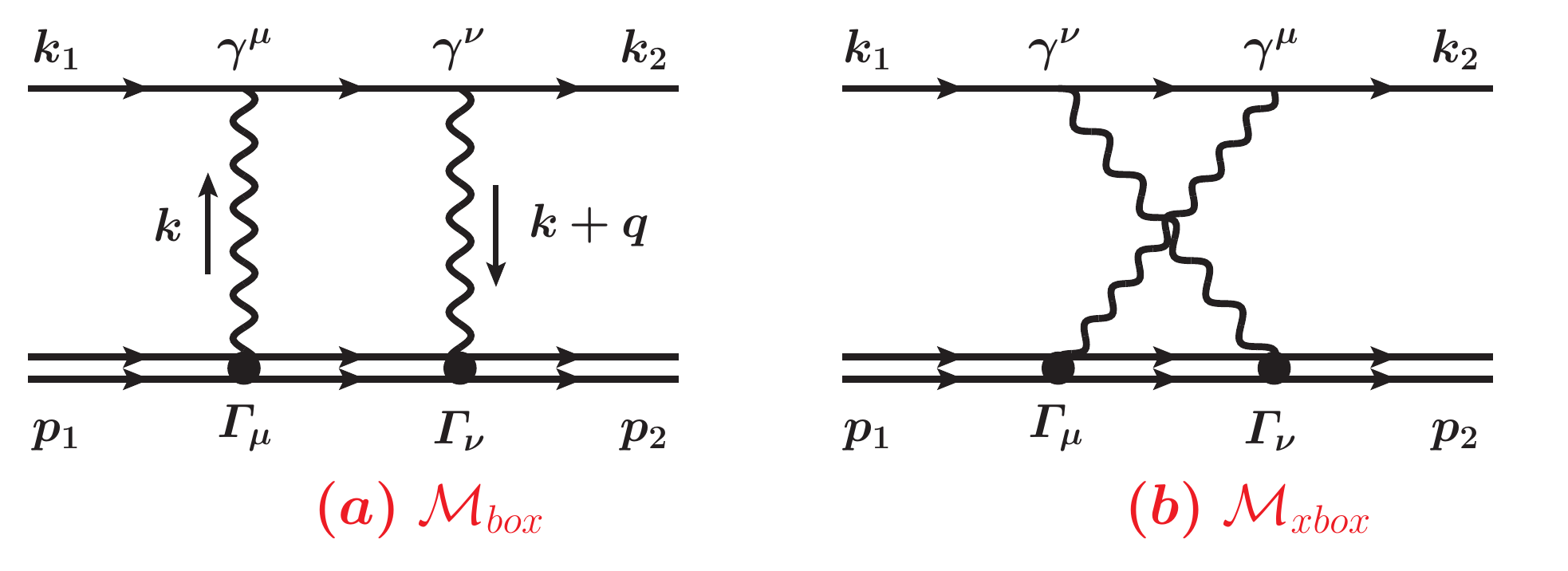}
    \caption{\label{fig:2}TPE box and crossed box diagrams}
\end{figure}
Feynman diagrams that represent TPE processes are shown in Fig. \ref{fig:2}. The corresponding amplitudes are given by
\begin{equation}\label{1.13}
\begin{split}
    M_{box} = & \int \frac{d^4k}{(2 \pi)^4} \ \frac{(-i)}{k^2 - \lambda^2} \frac{(-i)}{(k+q)^2 - \lambda^2} \\
    & \cdot \bar{u}(k_2) (z i e\gamma^\nu) \frac{i(\slashed{k}_1 + \slashed{k} + m)}{(k_1 + k)^2 - m^2} (z i e \gamma^\mu) u(k_1) \\
    & \cdot \bar{U}(p_2) (-i e \Gamma_\nu) \frac{i(\slashed{p}_1 - \slashed{k} + M)}{(p_1 - k)^2 - M^2} (-i e \Gamma_\mu) U(p_1), \\
    M_{xbox} = & \int \frac{d^4k}{(2 \pi)^4} \ \frac{(-i)}{k^2 - \lambda^2} \frac{(-i)}{(k+q)^2 - \lambda^2} \\
    & \cdot \bar{u}(k_2) (z i e \gamma^\mu) \frac{i(\slashed{k}_2 - \slashed{k} + m)}{(k_2 - k)^2 - m^2} (z i e \gamma^\nu) u(k_1) \\ & \cdot \bar{U}(p_2) (-i e \Gamma_\nu) \frac{i(\slashed{p}_1 - \slashed{k} + M)}{(p_1 - k)^2 - M^2} (-i e \Gamma_\mu) U(p_1) .
\end{split}
\end{equation}

In our approach, the model-independent TPE contribution can be obtained by imposing two assumptions on the matrix elements Eq. (\ref{1.13}). The first assumption is that one of the virtual photons in Fig. \ref{fig:2} is soft, i.e. it is transferring a vanishingly small momentum. The respective algebraical approximation has to be applied in the numerators and denominators of Eq. (\ref{1.13}). The second assumption is that the off-shell proton vertices $\Gamma_\mu$ and $\Gamma_\nu$ are given by their on-shell expression Eq. (\ref{1.9}). Based on these assumptions, a soft proton vertex $\Gamma_\mu(0)$ can be replaced by $\gamma_\mu$. With this in mind, let us now focus on the box diagram Fig. \ref{fig:2}(a). We denote $M_{box}^\prime$ to be the amplitude that corresponds to $k \rightarrow 0$ and $M_{box}^{\prime \prime}$ to be the amplitude that corresponds to $k+q \rightarrow 0$. As a result, one may find
\begin{equation}\label{1.14}
    M_{box}^\prime = - z \frac{\alpha}{2 \pi} b_{11} C_0 \Big[ (-k_1, m), (p_1, M) \Big] M_{1 \gamma},
\end{equation}
\begin{equation}\label{1.15}
    M_{box}^{\prime \prime} = - z \frac{\alpha}{2 \pi} b_{22} C_0 \Big[ (-k_2, m), (p_2, M) \Big] M_{1 \gamma},
\end{equation}
where $C_0 \Big[ (k_i, m), (p_j, M) \Big]$ is the Passarino-Veltman scalar three-point function defined as
\begin{equation}\label{1.16}
\begin{split}
   C_0 & \Big[ (k_i, m), (p_j, M) \Big] \\
   & \equiv \int \frac{d^4k}{i \pi^2} \ \frac{1}{[k^2-\lambda^2] [(k-k_i)^2 - m^2] [(k-p_j)^2 - M^2]}.
\end{split}
\end{equation}
This means that $M_{box} = M_{box}^\prime + M_{box}^{\prime \prime}$ can be written as
\begin{multline}\label{1.17}
    M_{box} = - z \frac{\alpha}{2 \pi} \Big( b_{11} C_0 \Big[ (-k_1, m), (p_1, M) \Big] \\
            + b_{22} C_0 \Big[ (-k_2, m), (p_2, M) \Big] \Big) M_{1 \gamma}.
\end{multline}
In a similar fashion, one can find the matrix element of the crossed box diagram Fig. \ref{fig:2}(b),
\begin{multline}\label{1.18}
    M_{xbox} = - z \frac{\alpha}{2 \pi} \Big( b_{12} C_0 \Big[ (k_1, m), (p_2, M) \Big] \\
            + b_{21} C_0 \Big[ (k_2, m), (p_1, M) \Big] \Big) M_{1 \gamma}.
\end{multline}
The analytical expressions for the scalar three-point functions that appear in Eqs. (\ref{1.17}) and (\ref{1.18}) are provided in Appendix \ref{1.39} (corresponding ultrarelativistic limit formulas can be found, e.g., in Ref. \cite{PhysRevD.88.053008}). It can be easily seen from those expressions that
\begin{equation*}
\begin{split}
    C_0 \Big[ (- k_1, m), (p_1, M) \Big] = & \ C_0 \Big[ (- k_2, m), (p_2, M) \Big], \\
    C_0 \Big[ (k_1, m), (p_2, M) \Big] = & \ C_0 \Big[ (k_2, m), (p_1, M) \Big].
\end{split}
\end{equation*}
Moreover, we remember that the condition Eq. (\ref{1.6}) holds true for the two-photon exchange processes shown in Fig. \ref{fig:2}. That is why
\begin{multline}\label{1.19}
    M_{2 \gamma} = M_{box} + M_{xbox} \ \ \ \ \\
                 = - z \frac{\alpha}{2 \pi} \bigg( 2 \ b_{11} C_0 \Big[ (-k_1, m), (p_1, M) \Big] \ \ \ \ \ \ \ \ \ \\
                 + 2 \ b_{12} C_0 \Big[(k_1, m), (p_2, M) \Big] \bigg) M_{1 \gamma}.
\end{multline}
It should be mentioned here that the imaginary part of the two-photon exchange amplitude $M_{2 \gamma}$ is contained solely in the box diagram, whereas the crossed box diagram is purely real.

Using the definition of Eq. (\ref{1.11}), we find the corresponding TPE correction to look like
\begin{multline}\label{1.20}
    \delta_{2 \gamma} = \frac{\alpha}{\pi} \mathrm{Re} \bigg( 2 \ b_{11} C_0 \Big[ (-k_1, m), (p_1, M) \Big] \\
    + 2 \ b_{12} C_0 \Big[(k_1, m), (p_2, M) \Big] \bigg).
\end{multline}
The obtained result Eq. (\ref{1.20}) is exact and we did not use an approximation $\mathrm{Re} \Big( C_0 \Big[ (- k_1, m), (p_1, M) \Big ] \Big ) = - C_0 \Big[ (k_1, m), (p_1, M) \Big]$ made by Tsai.

\section{Soft bremsstrahlung contributions}\label{1.46}

\begin{figure}[htp]
    \includegraphics[scale=0.45]{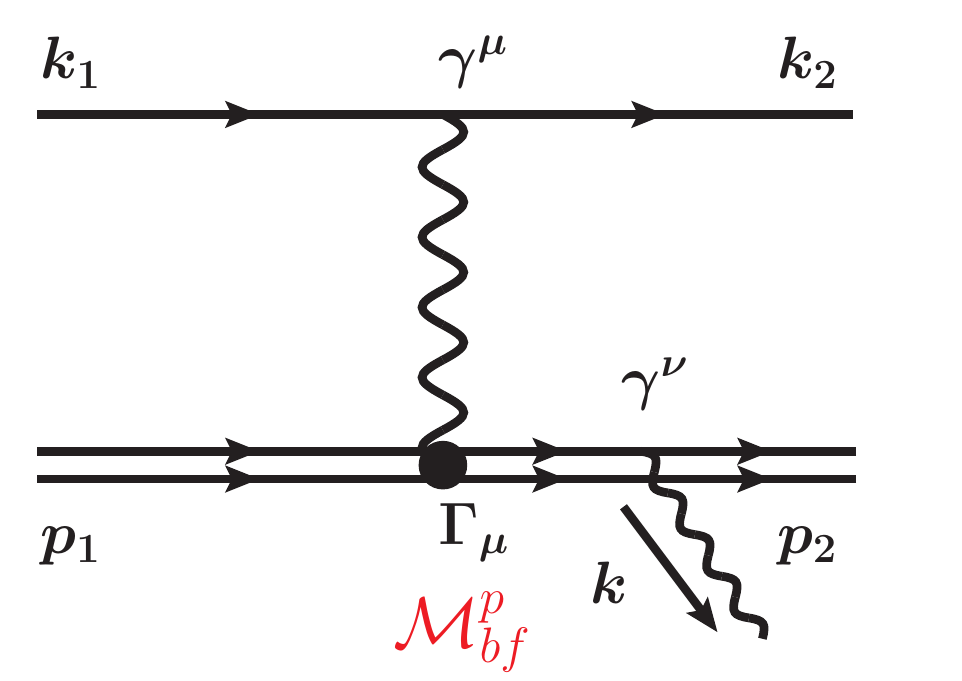}
    \caption{\label{fig:3}Proton leg bremsstrahlung diagram}
\end{figure}

Bremsstrahlung diagrams Figs. \ref{fig:1}(d)-\ref{fig:1}(g) are an inevitable part of charge asymmetry calculations. As it was pointed out earlier, one needs to consider the bremsstrahlung interference contribution to cancel out the infrared divergence of Eq. (\ref{1.20}).

In our approach, similarly to the TPE case, we assume that a vertex $\Gamma_\nu$ that depicts the coupling of a real soft photon to the proton Figs. \ref{fig:1}(f) and \ref{fig:1}(g) can be replaced by $\gamma_\nu$. In addition, we neglect the 4-momentum $k$ of the corresponding photon. As a result, the expressions for all four soft-bremsstrahlung matrix elements become alike and can be factorized by the Born amplitude Eq. (\ref{1.8}). For example, the matrix element shown in Fig. \ref{fig:3} takes the following form,
\begin{equation}\label{1.21}
\begin{split}
    M^p_{bf} = \ & \bar{U}(p_2) (- i e \gamma^\nu) \frac{i (\slashed{p}_2 + \slashed{k} + M)}{(p_2 + k)^2 - M^2}  (- i e \Gamma_\mu) U(p_1) \epsilon^*_\nu(k) \\
    \cdot \ & \bar{u}(k_2) (z i e \gamma^\mu) u(k_1) \frac{ (- i)}{q^2 - \lambda^2} = e \ \frac{(p_2 \cdot \epsilon^*)}{(p_2 \cdot k)} M_{1 \gamma},
\end{split}
\end{equation}
where $\epsilon^*_\nu(k)$ describes the polarization of the emitted photon.

The rest of the graphs in Figs. \ref{fig:1}(d)-\ref{fig:1}(g) can be found in a similar way, and the resulting bremsstrahlung amplitude can be written as
\begin{multline}\label{1.22}
    M_{b} = \Bigg[ z \bigg( \frac{(k_1 \cdot \epsilon^*)}{(k_1 \cdot k)} - \frac{(k_2 \cdot \epsilon^*)}{(k_2 \cdot k)} \bigg) \\
          + \bigg( \frac{(p_2 \cdot \epsilon^*)}{(p_2 \cdot k)} - \frac{(p_1 \cdot \epsilon^*)}{(p_1 \cdot k)} \bigg) \Bigg] e M_{1 \gamma}.
\end{multline}
By taking the square of Eq. (\ref{1.22}), summing over all polarization states, and factoring out the Born contribution Eq. \ref{1.12}, one may find the cross section for the bremsstrahlung process to look like
\begin{equation}\label{1.23}
    d \sigma_{b} = - \frac{\alpha}{4 \pi^2} \Big[ z^2 a_\Sigma - z b_\Sigma + c_\Sigma \Big] d \sigma_{1 \gamma},
\end{equation}
where
\begin{equation}\label{1.24}
\begin{split}
    a_\Sigma = & \sum_{i,j = 1,2}^{} a_{ij} L_{k_i, k_j} \theta(k_i) \theta(k_j), \ \ a_{ij} = (k_i \cdot k_j), \\
    b_\Sigma = & \sum_{i,j = 1,2}^{} b_{ij} L_{k_i, p_j} \theta(k_i) \theta(p_j), \ \ b_{ij} = 2 (k_i \cdot p_j), \\
    c_\Sigma = & \sum_{i,j = 1,2}^{} c_{ij} L_{p_i, p_j} \theta(p_i) \theta(p_j), \ \ c_{ij} = (p_i \cdot p_j), \\
    \theta(k_1) & =  \theta(p_1) \equiv + 1, \ \theta(k_2) = \theta(p_2) \equiv - 1
\end{split}
\end{equation}
and
\begin{equation}\label{1.25}
    L_{k_i, p_j} \equiv L_{ij} = \int \frac{d^3 \vec{k}}{\omega} \frac{1}{(k_i \cdot k)} \frac{1}{(p_j \cdot k)}.
\end{equation}
The expressions for $L_{k_i, k_j}$ and $L_{p_i, p_j}$ are defined identically to Eq. (\ref{1.25}); one just needs to replace there $p_j \rightarrow k_j$ and $k_i \rightarrow p_i$, correspondingly. The integral in  Eq. (\ref{1.25}) is infrared divergent. Similarly to the TPE case, this divergence can be regularized by assigning the mass $\lambda$ to the photon, so that $\omega = \sqrt{\vec{k}^2 + \lambda^2}$. After that, the integration over the emitted photon's phase space can be performed in the $R$ frame (sometimes called the $S^0$ frame) to avoid in Eq. (\ref{1.25}) the dependence on the angle at which the photon is emitted. The corresponding calculation has been performed by 't Hooft and Veltman \cite{THOOFT1979365}, and their result was summarized and rewritten in Ref. \cite{PhysRevC.62.054320} using the metric identical to ours.

In our calculations, we will mainly follow the derivation of Ref. \cite{THOOFT1979365} but single out several Passarino-Veltman three-point functions at final stages by using the following identity:
\begin{equation}\label{1.26}
    \int \limits_{0}^{1} \frac{dx}{u^2} \ln \Big( \frac{4 \Delta \varepsilon^2}{\lambda^2} \Big) = \int \limits_{0}^{1} \frac{dx}{u^2} \ln \Big( \frac{4 \Delta \varepsilon^2}{u^2} \Big) + \int \limits_{0}^{1} \frac{dx}{u^2} \ln \Big(\frac{u^2}{\lambda^2} \Big).
\end{equation}
Such an approach (mentioned, e.g., in Ref. \cite{0954-3899-41-11-115001}) enables us to completely absorb the infrared $\lambda$-dependence in three-point functions. Given that TPE contributions are also expressed in terms of infrared-divergent three-point functions Eq. (\ref{1.20}), it can be shown that the corresponding sum is infrared free.

Using the identity Eq. (\ref{1.26}), the result Eq. (4.13) of Ref. \cite{PhysRevC.62.054320} can be written as
\begin{equation}\label{1.27}
    L_{ij} = \frac{2 \pi}{\gamma_{ij}} \Big[ S^{(1)}_{ij} + S^{(2)}_{ij}\Big],
\end{equation}
where, one can find a modified form for $S^{(1)}_{ij}$ to look like
\begin{equation}\label{1.28}
\begin{split}
    S^{(1)}_{ij} = \ & \frac{2 \alpha_{ij} \gamma_{ij}}{\alpha^2_{ij} m^2 - M^2} \Bigg[ \ln^2 \Big( \frac{2 \Delta \epsilon}{M} \Big) - \ln^2 \Big( \frac{2 \Delta \epsilon}{\alpha_{ij} m} \Big) \Bigg] \\
                 - \ & 2 \alpha_{ij} C_0 \Big[(\alpha_{ij} k_i, \alpha_{ij} m), (p_j, M) \Big]
\end{split}
\end{equation}
with
\begin{equation}\label{1.29}
\begin{split}
    \alpha_{ij} \equiv \ & \frac{k_i \cdot p_j + \gamma_{ij}}{m^2} = \frac{b_{ij} + 2 \gamma_{ij}}{2 m^2}, \\
    \gamma_{ij} \equiv \ & \sqrt{(k_i \cdot p_j)^2 - m^2 M^2} = \frac{1}{2} \sqrt{b^2_{ij} - 4 m^2 M^2}. \\
\end{split}
\end{equation}
The second term of Eq. (\ref{1.27}) contains dilogarithm functions $Li_2$ that have a cut along the positive real axis starting at $x = 1$. This term is infrared free and does not require any modifications
\begin{equation}\label{1.30}
\begin{split}
    S^{(2)}_{ij} = & \ln^2 \Big( \frac{\beta_{i}}{m M} \Big) - \ln^2 \Big( \frac{\delta_{j}}{M^2} \Big) \\
    + & \mathrm{Li_2} \Big( 1 - \frac{\beta_{i} (l_{ij} \cdot p_2)}{\gamma_{ij} M^2} \Big) + \mathrm{Li_2} \Big( 1 - \frac{m^2 (l_{ij} \cdot p_2)}{\gamma_{ij} \beta_{i}} \Big) \\
    - & \mathrm{Li_2} \Big( 1 - \frac{\delta_{j} (l_{ij} \cdot p_2)}{M^2 \alpha_{ij} \gamma_{ij}} \Big) - \mathrm{Li_2} \Big( 1 - \frac{M^2 (l_{ij} \cdot p_2)}{\alpha_{ij} \gamma_{ij} \delta_{j}} \Big)
\end{split}
\end{equation}
with
\begin{equation}\label{1.31}
\begin{split}
        \beta_i \equiv \ & (k_i \cdot t) + \sqrt{(k_i \cdot t)^2 - m^2 t^2}, \\
       \delta_j \equiv \ & (p_j \cdot t) + \sqrt{(p_j \cdot t)^2 - M^2 t^2}, \\
         l_{ij} \equiv \ & \alpha_{ij} k_i, \\
              t \equiv \ & p_2 + k = p_1 + k_1 - k_2.
\end{split}
\end{equation}
At this point, it is worth mentioning that $\Delta \epsilon$ is the upper limit of the integration over the photon energy in Eq. (\ref{1.25}). This quantity is chosen to be smaller than any of the other energies of an experiment and serves as a parameter that splits the photon's phase space into the soft and hard regions. The relevant discussions can be found in Refs. \cite{RevModPhys.41.193, PhysRevC.62.054320}. Here, we just note that $\Delta \epsilon$ is defined in the $R$ frame and one usually wants to be able to relate this quantity to some energy scale that describes an experimental setup. For example, in Ref. \cite{PhysRevC.62.054320}, the authors show how to relate $\Delta \epsilon$ to the final electron detector acceptance. In our work, we chose to relate $\Delta \epsilon$ to the frame-invariant quantity that is called inelasticity and defined as
\begin{equation}\label{1.32}
    \nu \equiv (p_2 + k)^2 - M^2 = t^2 - M^2.
\end{equation}
By using this definition and following the discussion on $\Delta \epsilon$ given in Ref. \cite{PhysRevC.62.054320}, one may find
\begin{equation}\label{1.33}
    \nu = 2 \sqrt{t^2} \Delta \epsilon \approx 2 M \Delta \epsilon,
\end{equation}
where the last equity is achieved by imposing the soft-photon approximation $k \rightarrow 0$.

The result of Eq. (\ref{1.33}) enables us to combine our soft-photon calculations with the hard-photon computations that follow the procedure developed by Bardin and Shumeiko \cite{BARDIN1977242}, where the authors use the definition for inelasticity identical to ours. It should be noted here that the approach of Ref. \cite{BARDIN1977242} is advantageous over the approach of Refs. \cite{PhysRev.122.1898, RevModPhys.41.205}, because the final result for the total radiative correction (the sum of soft and hard contributions) appears to be independent of the artificial parameter that separates the photon's phase space into soft and hard regions. In our paper, we perform the model-independent part of the total calculation and just mention that, due to complicated detector geometry, the corresponding model-dependent hard-photon computations are performed numerically using Monte Carlo codes (see, e.g., Refs. \cite{Akushevich20121448, Akushevich2015}).

Let us comment on the inelasticity. This quantity, which describes a $2 \rightarrow 3$ process Eq. (\ref{1.2}), is limited by the respective kinematic bound $\nu_{\mathrm{\max}}$ that looks like \cite{PhysRevD.64.113009}
\begin{equation}\label{1.34}
    \nu_{\mathrm{max}} = \frac{1}{2 m^2} \bigg( 2 \gamma_{11} \sqrt{Q^2(Q^2 + 4m^2)} - Q^2 (m^2 + b_{11} ) \bigg).
\end{equation}
Besides the kinematic bound of Eq. (\ref{1.34}), the integration over the photon's phase space in Eq. (\ref{1.25}) can be limited by the properties of a detector system that sets the corresponding cut value $\nu_{\mathrm{cut}}$ on the inelasticity. For example, this cut value can be set below the pion production threshold ($\nu_{\mathrm{cut}} \simeq 0.271$). Another possible choice for the cut can be dictated by the inability of the detector to register scattered leptons of small energies. In this case, one usually wants to relate $\varepsilon_2$ to $\nu_{\mathrm{cut}}$. The corresponding lab frame relation can be obtained from Eq. (\ref{1.32}) and is given by
\begin{equation}\label{1.35}
    \nu_{\mathrm{cut}} = b_{11} - Q^2 - 2 m \varepsilon_2.
\end{equation}
As a result, in the presence of experimental cuts, the upper integration limit in Eq. (\ref{1.25}) has to be chosen based on the following condition
\begin{equation}\label{1.36}
    \nu = \mathrm{Min} \ (\nu_{\mathrm{cut}}, \ \nu_{\mathrm{max}}).
\end{equation}

At this point, we can write down an invariant form for the charge-dependent soft-bremsstrahlung correction to the cross section in accordance with definition of Eqs. (\ref{1.12}) and (\ref{1.23}):

\begin{multline}\label{1.37}
               \delta_{b} = - \frac{\alpha}{\pi} \mathrm{Re} \Bigg( \frac{b_{11}}{2 \gamma_{11}} \Big( S^{(2)}_{11} + S^{(2)}_{22} \Big) - \frac{b_{12}}{2 \gamma_{12}} \Big( S^{(2)}_{12} + S^{(2)}_{21} \Big) \\
               \ \ \ \ \ \ \ \ \ \ \ + \frac{2 \alpha_{11} b_{11}}{\alpha^2_{11} m^2 - M^2} \bigg[ \ln^2 \Big(\frac {\nu}{M^2} \Big) - \ln^2 \Big( \frac{\nu}{\alpha_{11} m M} \Big) \bigg] \\
               \ \ \ \ \ \ \ \ \ \ \ - \frac{2 \alpha_{12} b_{12}}{\alpha^2_{12} m^2 - M^2} \bigg[ \ln^2 \Big( \frac{\nu}{M^2} \Big) - \ln^2 \Big( \frac{\nu}{\alpha_{12} m M} \Big) \bigg] \\
              \ \ - 2 \ b_{11} \alpha_{11} C_0 \Big[ (\alpha_{11} k_1, \alpha_{11} m), (p_1, M) \Big] \\
              \ + 2 \ b_{12} \alpha_{12} C_0 \Big[ (\alpha_{12} k_1, \alpha_{12} m), (p_2, M) \Big] \bigg).
\end{multline}
As intended, the $\lambda$-dependence is now solely ``hidden'' in the last two terms of Eq. (\ref{1.37}) that are given by scalar three-point functions, the exact expressions for which are provided in Appendix \ref{1.39}. By assigning the fictitious mass to the photon and by choosing $m \ll \varepsilon_1$ we confirmed that the predictions of Eq. (\ref{1.37}) are identical to those of the Ref. \cite{PhysRevC.62.054320}.

\section{Results}\label{1.47}

The results given in Eqs. (\ref{1.20}) and (\ref{1.37}) may be added together to give us the charge asymmetry in the soft-photon approximation. It should be mentioned here that the direct sum of these expressions has to be shifted by a constant factor to provide a physically justified asymmetry that implies zero asymmetry at $Q^2 = 0$. This asymmetry is found to be
\begin{widetext}
\begin{equation}\label{1.38}
\begin{split}
    \delta = - \frac{\alpha}{\pi} \Bigg( \frac{b_{12}}{\gamma_{12}} \ & \bigg[ \frac{1}{2} \ln \Big( \alpha_{12} \Big) \cdot \ln \Big(  \frac{4 \gamma^2_{12}}{ m^4 \alpha_{12} (1 - \alpha_{12})^2} \Big) + \mathrm{Li_2} \Big( \frac{u}{2 \gamma_{12} (1 -  \alpha_{12})} \Big) - \mathrm{Li_2} \Big( \frac{ u \alpha_{12}}{2 \gamma_{12} (1 -  \alpha_{12})} \Big) \bigg] \\
    - \frac{b_{11}}{\gamma_{11}} \ & \bigg[ \frac{1}{2} \ln \Big( \alpha_{11} \Big) \cdot \ln \Big(  \frac{4 \gamma^2_{11}}{ m^4 \alpha_{11} (1 - \alpha_{11})^2} \Big) + \mathrm{Li_2} \Big( \frac{2 m^2 + 2 M^2 - s}{2 \gamma_{11} (1 -  \alpha_{11})} \Big) - \mathrm{Li_2} \Big( \frac{(2 m^2 + 2 M^2 - s) \alpha_{11}}{2 \gamma_{11} (1 -  \alpha_{11})} \Big) \bigg] \\
    + & \frac{2 \alpha_{11} b_{11}}{\alpha^2_{11} m^2 - M^2} \bigg[ \ln^2 \Big( \frac{\nu}{M^2} \Big) - \ln^2 \Big( \frac{\nu}{\alpha_{11} m M} \Big) \bigg] - \frac{2 \alpha_{12} b_{12}}{\alpha^2_{12} m^2 - M^2} \bigg[ \ln^2 \Big( \frac{\nu}{M^2} \Big) - \ln^2 \Big( \frac{\nu}{\alpha_{12} m M} \Big) \bigg] \\
    + & \frac{b_{11}}{2 \gamma_{11}} \bigg[ S^{(2)}_{11} + S^{(2)}_{22} \bigg] - \frac{b_{12}}{2 \gamma_{12}} \bigg[ S^{(2)}_{12} + S^{(2)}_{21} \bigg] \Bigg).
\end{split}
\end{equation}
\end{widetext}

\begin{figure*}[t]
    \includegraphics[scale=0.45]{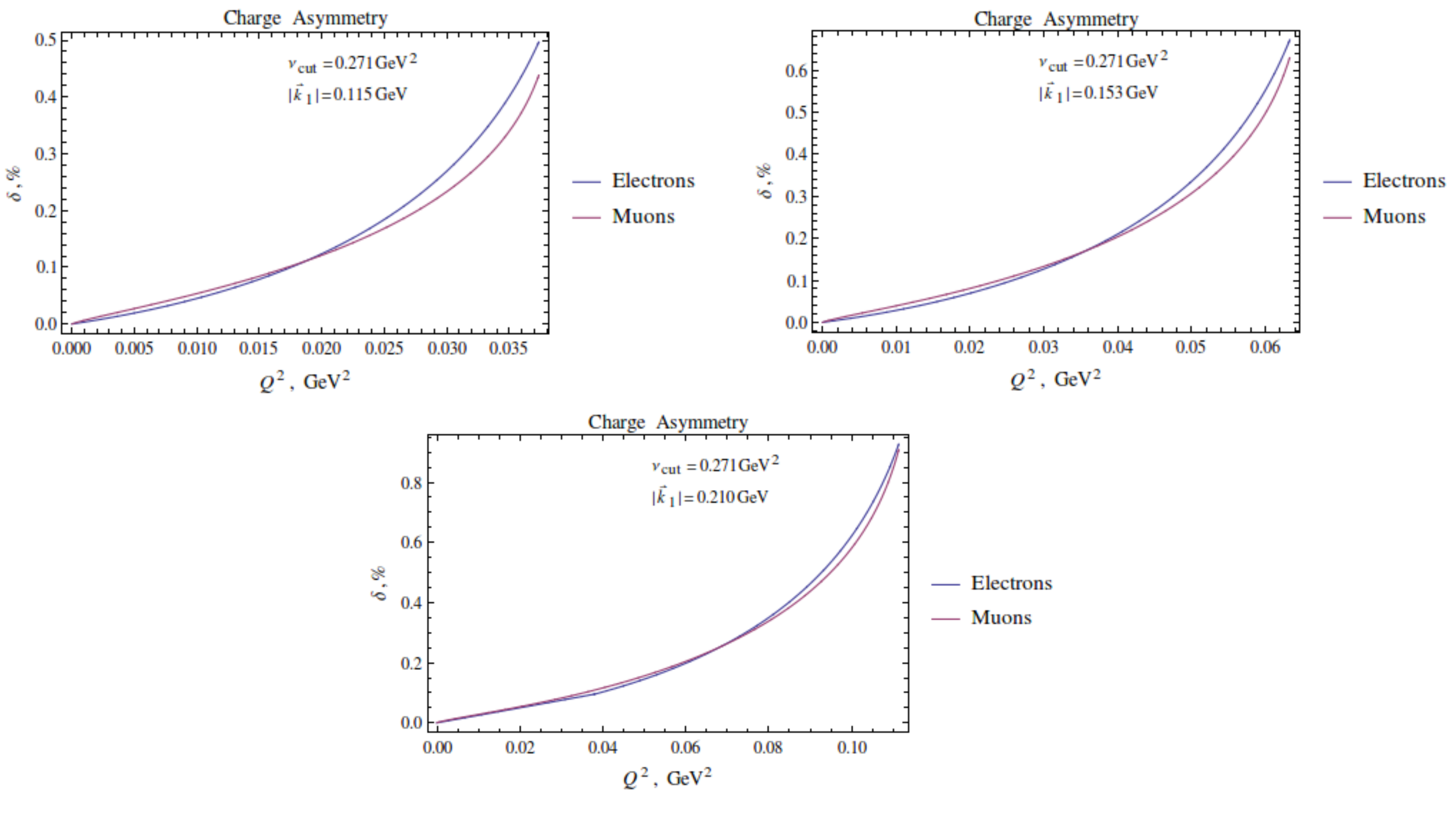}
    \caption{\label{fig:4}Charge asymmetry predictions for MUSE}
\end{figure*}

In the limit $Q^2 \rightarrow 0$ the variables in Eq. (\ref{1.38}) behave as follows: $\nu \rightarrow 0, \ b_{12} \rightarrow b_{11}, \ \gamma_{12} \rightarrow \gamma_{11}, \ \alpha_{12} \rightarrow \alpha_{11}$. This assures that the terms that contain singularities at $Q^2 \rightarrow 0$ cancel each other out.

The obtained result Eq. (\ref{1.38}) can be used to provide model-independent predictions for any lepton-proton scattering experiment that requires lepton mass effects to be taken into account. In particular, the respective predictions in the kinematics of MUSE are shown in Fig. \ref{fig:4}. The electron-positron charge asymmetry there represents the ultrarelativistic (massless) prediction, whereas the effects of the lepton's mass can be seen on the graph for the muon-antimuon asymmetry.

\section{Conclusion}\label{1.48}

We have calculated analytically the charge asymmetry in elastic lepton-proton scattering to order $\alpha^3$ by generalizing the soft-photon approach of Tsai. The present calculation improves the result of Tsai in three essential aspects. First, we do not neglect the mass of the lepton as compared to its energy. Second, we do not use the approximation $\mathrm{Re} \Big( C_0 \Big[ (- k_1, m), (p_1, M) \Big] \Big ) = - C_0 \Big[ (k_1, m), (p_1, M) \Big]$ in an evaluation of soft TPE. The corresponding exact expression for $C_0 \Big[ (- k_1, m), (p_1, M) \Big]$ is given in Appendix \ref{1.39}. Finally, the charge-dependent contribution from the emission of real soft photons is calculated exactly without the approximation made by Tsai.

In the ultrarelativistic limit, if one corrects the result of Tsai by the terms left out by his approximations, our predictions coincide with those of Ref. \cite{PhysRev.122.1898}. The corresponding ultrarelativistic analysis of a difference between exact expressions and expressions of Tsai was done in Ref. \cite{Gerasimov2015}. It should also be mentioned here that unlike Tsai,
in the bremsstrahlung contribution calculation, we isolate three-point functions that have a form different from those three-point functions that are coming from the TPE contribution. In other words, in our derivations three-point functions coming from the TPE contribution and three-point functions coming from the bremsstrahlung contribution do not cancel each other out. However, as expected, the sum of the bremsstrahlung and TPE contributions stays independent of the chosen approach.

The obtained results in the kinematics of MUSE can be compared to the TPE calculation of Tomalak and Vanderhaeghen \cite{PhysRevD.90.013006} that was also performed without neglecting the mass of the lepton. In addition, this calculation provides more confidence on the fact that soft photon TPE and bremsstrahlung corrections that appear to be on the order of a few tenth percent in the kinematics of MUSE are about an order of magnitude larger than helicity-flip contributions for muons, which were recently estimated in Refs. \cite{PhysRevD.94.116007, PhysRevC.95.025203}.

Our analytical results can be combined with hard photon calculations that take into account the mass of the lepton to provide complete radiative corrections predictions for elastic lepton-proton scattering experiments.

\begin{acknowledgments}
We thank W. J. Briscoe, E. J. Downie, A. Gramolin, I. Lavrukhin and M. Mai for useful discussions. This work was supported in part by The George Washington University through the Gus Weiss endowment and in part by a JSA/JLab Graduate Fellowship Award.
\end{acknowledgments}

\appendix
\begin{widetext}

\section{Scalar three-point functions}\label{1.39}

Using the approach described in Ref. \cite{THOOFT1979365} one may derive the following expression for the scalar 3-point functions defined in Eq. (\ref{1.16})
\begin{multline}\label{1.40}
     C_0 \Big[ (k_i, m), (p_j, M) \Big] = - \frac{1}{2 \gamma_{ij}} \Bigg[\ln \Big( \frac{\lambda^2}{m M} \Big)  \ln \Big( \frac{M}{m \alpha_{ij}} \Big) + \frac{1}{2} \ln \Big( \alpha_{ij} \Big) \ln \Big( \frac{4 \gamma^2_{ij} \alpha_{ij}}{m^2 M^2 (1 - \alpha_{ij})^2} \Big) \\
     - \mathrm{Li_2} \Big( \frac{\alpha_{ij} (m^2 - b_{ij} + M^2)}{2 \gamma_{ij} (1 - \alpha_{ij})} \Big) + \mathrm{Li_2} \Big( \frac{m^2 - b_{ij} + M^2}{2 \gamma_{ij} (1 - \alpha_{ij})} \Big) \Bigg]
\end{multline}
The obtained result for $C_0 \Big[ (k_i, m), (p_j, M) \Big]$ is real for any combination of $k_i = k_1, \ k_2$ and $p_j = p_1, \ p_2$. The predictions of Eq. (\ref{1.40}) were checked numerically in the kinematics of MUSE using the LOOPTOOLS \cite{HAHN1999153} package.

Using the result of Eq. (\ref{1.40}), three-point functions that are coming from the bremsstrahlung contribution can be found to have the following simple form:
\begin{equation}\label{1.41}
\begin{split}
     C_0 \Big[ (\alpha_{11} k_1, \alpha_{11} m), (p_1, M) \Big] =& - \frac{1}{2 \alpha_{11} \gamma_{11}}\ln \Big( \frac{\lambda^2}{\alpha_{11} m M} \Big)  \ln \Big( \frac{M}{\alpha_{11} m} \Big), \\
     C_0 \Big[ (\alpha_{12} k_1, \alpha_{12} m), (p_2, M) \Big] =& - \frac{1}{2 \alpha_{12} \gamma_{12}}\ln \Big( \frac{\lambda^2}{\alpha_{12} m M} \Big)  \ln \Big( \frac{M}{\alpha_{12} m} \Big). \\
\end{split}
\end{equation}
Three-point functions that are coming from the TPE contribution are found to be
\begin{equation}\label{1.42}
\begin{split}
    C_0 \Big[ (k_1, m), (p_2, M) \Big] = - \frac{1}{2 \gamma_{12}} \Bigg[\ln \Big( \frac{\lambda^2}{m M} \Big)  \ln \Big( \frac{M}{m \alpha_{12}} \Big) + & \frac{1}{2} \ln \Big( \alpha_{12} \Big) \ln \Big( \frac{4 \gamma^2_{12} \alpha_{12}}{m^2 M^2 (1 - \alpha_{12})^2} \Big) \\
     -& \mathrm{Li_2} \Big( \frac{\alpha_{12} u}{2 \gamma_{12} (1 - \alpha_{12})} \Big) + \mathrm{Li_2} \Big( \frac{u}{2 \gamma_{12} (1 - \alpha_{12})} \Big) \Bigg], \\
     C_0 \Big[ (- k_1, m), (p_1, M) \Big] = - \frac{1}{2 \gamma_{11}} \Bigg[ \ln \Big( \frac{\lambda^2}{m M} \Big)  \ln \Big( \frac{M}{m \alpha^{'}_{11}} \Big) + & \frac{1}{2} \ln \Big( \alpha^{'}_{11} \Big) \ln \Big( \frac{4 \gamma^2_{11} \alpha^{'}_{11}}{m^2 M^2 (1 + \alpha^{'}_{11})^2} \Big) \\
     + \frac{5 \pi^2}{6} - \frac{1}{2} \ln^2 \Big( & \frac{s}{2 \gamma_{11} (1 + \alpha^{'}_{11})} \Big) + i \pi \ln \Big( \frac{s}{\lambda^2 (1 + \alpha^{'}_{11})^2}\Big) \\
     -& \mathrm{Li_2} \Big( \frac{- \alpha^{'}_{11} s}{2 \gamma_{11} (1 + \alpha^{'}_{11})} \Big) - \mathrm{Li_2} \Big( \frac{2 \gamma_{11} (1 + \alpha^{'}_{11})}{s} \Big),
      \Bigg]
\end{split}
\end{equation}
where
\begin{equation}\label{1.43}
\alpha^{'}_{11} \equiv - \alpha_{-11} = - \Big( \frac{- k_1 \cdot p_1 + \gamma_{11}}{m^2} \Big) = - \Big( \frac{- b_{11} + \gamma_{11}}{2 m^2} \Big) > 0.
\end{equation}

\end{widetext}

\bibliography{Asymmetry}

\end{document}